\documentclass[aps, prd, twocolumn, showpacs, superscriptaddress, groupedaddress]{revtex4-1} 
\usepackage{graphicx}  
\usepackage{subcaption}
\usepackage{float}
\usepackage{tabularx}
\usepackage{array}
\usepackage{amssymb}
\usepackage{mathtools}
\usepackage{dcolumn}
\usepackage{color}
\usepackage[pagebackref=false, colorlinks=true]{hyperref}
\hypersetup{
linkcolor=blue,     
citecolor=blue,     
urlcolor=blue}   
\usepackage{amsmath}
\usepackage{ulem}
\usepackage[utf8]{inputenc}

\begin{document}
\title{Gravitational collapse of scalar and vector fields}

\author{Karim Mosani}
\email{kmosani2014@gmail.com}
\affiliation{International Centre for Space and Cosmology, School of Arts and Sciences, Ahmedabad University, Ahmedabad-380009 (Guj), India.}

\author{Koushiki}
\email{koushiki.malda@gmail.com}
\affiliation{International Centre for Space and Cosmology, School of Arts and Sciences, Ahmedabad University, Ahmedabad-380009 (Guj), India.}

\author{Pankaj S. Joshi}
\email{pankaj.joshi@ahduni.edu.in}
\affiliation{International Centre for Space and Cosmology, School of Arts and Sciences, Ahmedabad University, Ahmedabad-380009 (Guj), India.}

\author{Jay Verma Trivedi}
\email{jay.verma2210@gmail.com}
\affiliation{International Centre for Space and Cosmology, School of Arts and Sciences, Ahmedabad University, Ahmedabad-380009 (Guj), India.}

\author{Tapobroto Bhanja}
\email{tapobroto.bhanja@gmail.com}
\affiliation{International Center for Cosmology, \& PDPIAS,\\ Charotar University of Science and Technology, Anand- 388421 (Guj), India}


\date{\today}
\begin{abstract}
We study here the unhindered gravitational collapse of spatially homogeneous (SH) scalar fields $\phi$ with a potential $V_{s}(\phi)$, as well as vector fields $\tilde{A}$ with a potential $V_{v}(B)$ where $B=g(\tilde{A},\tilde{A})$ and $g$ is the metric tensor.
We show that in both cases, classes of potentials exist that give rise to black holes or naked singularities depending on the choice of the  potential. The strength of the naked singularity is examined, and they are seen to be strong, in the sense of Tipler, for a wide class of respective potentials. We match the collapsing scalar/vector field with a generalized Vaidya spacetime outside. We highlight that full generality is maintained within the domain of SH scalar or vector field collapse.\\

\textbf{keywords}: Gravitational collapse, singularity, scalar field, vector field, causal structure.

\end{abstract}
\maketitle
\section{Introduction}
The contraction of a matter field under its gravitational influence is called gravitational collapse. In 1939, Oppenheimer and Snyder \cite{Oppenheimer1939}, and independently in 1938, Datt \cite{Datt} developed the first solution of Einstein's field equations (called the OSD model) depicting the gravitational collapse of a massive star. They considered a very specific case of spatially homogeneous (SH) dust collapse (By spatial homogeneity, we mean homogeneous on a three-dimensional spacelike orbit with a six-dimensional isometry group $G_6$ corresponding to the spacetime \cite{Wainwright_2005}). 
Such a matter field undergoes gravitational collapse that ends up in a \textit{singularity}. Such a spacetime singularity is hidden behind an \textit{event horizon}, not visible to any observer, and what we obtain is a \textit{black hole} as the outcome of continual collapse. 
%
%
%

Extending the above special scenario, in 1969, Penrose proposed what is now known as the cosmic censorship hypothesis (CCH) 
\cite{penrose2}. 
The weaker version of the hypothesis states that all singularities of gravitational collapse are hidden within a black hole and hence, cannot be seen by a distant observer (a globally naked singularity cannot exist). The strong version of the hypothesis states that no past inextendible nonspacelike geodesics can exist between the singularity and any point in the spacetime manifold. In other words, a causal geodesic with a positive tangent ``at" the singularity does not exist (a locally naked singularity also cannot exist). The supporting argument for the validity of the strong CCH is the desirability of the spacetime manifold to be globally hyperbolic. Global hyperbolicity implies the existence of Cauchy surfaces embedded in the total manifold, thereby making general relativity a deterministic theory 
\cite{Geroch_70, hawking, Joshi_Global}.

Now singularity theorems of Hawking and Penrose
\cite{hawking, penrose}
do not imply that singularities are hidden from an external observer under any possible circumstances. In fact, singularity theorems take the causality condition as one of the axioms to start with to prove the existence of incomplete past (future) directed causal curves. Additionally, the OSD model that motivated cosmic censorship is a special case. Joshi and Malafarina \cite{Joshi2011} showed that any arbitrarily small neighbourhood of the initial data giving rise to OSD collapse contains initial data corresponding to collapse evolution giving rise to a singularity with the following property: one could trace outgoing past singular causal geodesics. This means that the end state of OSD collapse is unstable under small perturbations in initial data. Moreover, one can show the formation of naked singularities (global and local) as an end state of gravitational collapse from suitable, physically reasonable initial data for various matter fields
\cite{psjoshi2, Mosani2020}. \textcolor{black}{This implies that the initial conditions must be fine-tuned for the cosmic censorship conjecture to hold.}

In such a context, an important question one can ask here is as follows: what will be the end state of an unhindered gravitational collapse of a fundamental matter field, such as a scalar field or a vector field, derived from an appropriate Lagrangian?

The answer to this question has been achieved up to a certain extent. Scalar fields are fundamental matter fields derived from suitable Lagrangian. A real scalar field is a map defined on a smooth manifold as $\phi: \mathcal{M} \to \mathbb{R}$ with a suitable continuity condition. 
Christodoulou showed that in the case of gravitational collapse of a massless scalar field $\phi$ (the scalar field Lagrangian is $\mathcal{L}_{\phi}=(1/2) g^{\mu \nu}\partial_{\mu}\phi \partial_{\nu} \phi$), the set of initial data giving rise to a naked singularity as an end state has positive codimension in the entire initial data set
\cite{Christodoulou_94, Christodoulou_99}. 
This means that the initial data set corresponding to naked singularity has a zero measure in the total initial data set. In other words, naked singularity in such cases is unstable under arbitrarily small  perturbations in the initial data. 

One can have a massless scalar field with a potential function $V_s(\phi)$ that still be a fundamental matter field. A massive scalar field will then be a particular case of a massless scalar field with a specific potential of the form $V_s(\phi)=(1/2)\mu^2 \phi^2$, where $\mu$ is the mass term. Goswami and Joshi
\cite{goswami2007}
showed the example of the gravitational collapse of a massless  SH scalar field with a certain potential $V_s(\phi)$ that ends up in a naked singularity. Mosani, Dey, Bhattacharya, and Joshi
\cite{Mosani2022}
conducted a similar investigation for a massless scalar field with a two-dimensional analogue of the Mexican hat-shaped Higgs field potential and found out that the end state of such unhindered scalar field collapse is a naked singularity. 

In addition to scalar fields as fundamental matter fields, vector fields are also fundamental matter fields derived from suitable matter Lagrangian. Geometrically, vector fields on a smooth manifold $\mathcal{M}$ can be thought of as sections on the tangent bundle $\pi: T\mathcal{M} \to \mathcal{M}$, where $\pi$ is a continuous surjection. A section is a smooth map $\sigma: \mathcal{M} \to T\mathcal{M}$ such that $\pi \circ \sigma$ is an identity map on $\mathcal{M}$. From a particle physics point of view,  the fundamental nature of a vector field is different from that of a scalar field. There are many aspects, but one of the most important ones is that massive or massless vector fields mediate most particle physics processes. These represent the three fundamental interactions: quantum electrodynamics and weak and strong processes. A massless vector field with a potential function $V_v(B)$ is again a fundamental matter field. A massive vector field will then be a particular case of a massless vector field with a specific potential of the form $V_v(B)=(1/2)\mu^2 B$, where $\mu$ is the mass term. Garfinkle, Mann, and Vuille \cite{Garfinkle} have studied the collapse of a massive vector field and numerically obtained the critical initial conditions. To our knowledge, much analytical work has not been done in investigating the causal structure of the end-state spacetime of the unhindered gravitational collapse of matter fields that are vector fields.

In this paper, in both the massless SH scalar field as well as vector field cases, we show that there are broad classes of potentials for which the configuration collapses and ends up in either a black hole or a naked singularity depending on the potential function chosen. We approach the causality investigation problem of scalar field as well as vector field collapse in a unified way, so to speak. As far as general relativity is concerned, it does not discriminate between whether a scalar field or a vector field seeds the matter field. The matter field is entirely identified by a rank two tensor field that we call the stress-energy tensor. As far as SH perfect fluid is concerned, one can identify a given matter field by the functional form of the equation of state parameter $\omega (a)$, where $a$ is the scale factor of the collapsing cloud. We derive relevant equations of collapsing SH scalar field $\phi(a)$ and vector field $\Tilde{A}(a)$ in the sub-sections of section II. The main body of section II contains discussions and relevant relations regarding the gravitational collapse of SH perfect fluids. In section III, we smoothly join the interior collapsing perfect fluid with an external generalized Vaidya spacetime. In section IV, we investigate the causal structure of the spacetime (condition of obtaining a naked singularity) at the end of the collapse of the interior perfect fluid that is either a scalar field $\phi$ with potential $V_s$ or a vector field $\Tilde{A}$ with potential $V_{v}$. We also depict a few examples of well-known scalar fields and vector fields. In section V, we derive the criteria for the singularity, thus obtained in the end, to be strong of Tipler's type. In the last section, we highlight the key points of the investigation. Here we use the geometrized units $8\pi G=c=1$ throughout.

\section{Interior Collapsing matter field}
Consider a gravitational collapse of a SH perfect fluid. The components of the stress-energy tensor in the coordinate basis $\{dx^{\mu} \bigotimes \partial_{\nu} \vert 0 \leq \mu, \nu \leq 3\}$ of the comoving coordinates $(t,x,y,z)$  are given by
\begin{equation}
    T^{\mu}_{\nu}=\textrm{diag}\left(-\rho,p,p,p\right).
\end{equation}
  The spacetime geometry is governed by the flat ($k=0$) Friedmann–Lemaître–Robertson–Walker (FLRW) metric
\begin{equation}\label{FLRW1}
  ds^2=-dt^2+ a^2 d\Sigma^2,
\end{equation}
where $d\Sigma^2=dx^2 + dy^2+dz^2$. Here $a=a(t)$ is the scale factor such that $a(0)=1$ and $a(t_s)=0$, where $t_s$ is the time of formation of the singularity. $R=R(t,r)$ is the physical radius of the collapsing cloud and can be written as
\begin{equation}
    R(t,r)=r a(t),
\end{equation}
where $r$ is the radial spherical coordinate.
For a FLRW spacetime Eq.(\ref{FLRW1}), we have
\begin{equation}\label{rhoF}
    \rho=\frac{3\dot a^2}{a^2},
\end{equation}
and
\begin{equation}\label{pF}
    p=-\frac{2\ddot a}{a}-\frac{\dot a^2}{a^2}.
\end{equation}
The overhead dot denotes the partial time derivative of $a$. Eq.(\ref{rhoF}) can be rewritten to obtain the dynamics of the collapse as
\begin{equation}\label{adot}
    \dot a=-\sqrt{\frac{\rho(a)}{3}}a.
\end{equation}
Differentiating the above equation once again gives us
\begin{equation}\label{ddota}
    \ddot a =\frac{1}{3}a\left(\frac{a \rho_{,a}}{2}+\rho\right).
\end{equation}
Integrating Eq.(\ref{adot}), we obtain the time curve, which is
\begin{equation}
    t(a)=\int^{1}_{a}\sqrt{\frac{3}{\rho}}\frac{da}{a}.
\end{equation}
The dynamics of the scale factor $a(t)$ is, thus, the inverse of the LHS of the above equation. The time of formation of the singularity $t_s=t(0)$ is
\begin{equation}\label{times}
    t_s=\int^{1}_{0}\sqrt{\frac{3}{\rho}}\frac{da}{a}.
\end{equation}
Now, let us consider a particular matter field $\hat{T}$ from a set of all the possible SH perfect fluids. Choosing such an element means choosing a specific functional form of the equation of state parameter 
\begin{equation}\label{eosp}
    \omega (a)=\frac{p}{\rho}.
\end{equation} 
Using Eq.(\ref{rhoF}), Eq.(\ref{pF}), and Eq.(\ref{eosp}), we can express the density of the matter field with the equation of state parameter $\omega$ as
\begin{equation}\label{rhoomega}
    \rho=\rho_0\exp\left( \int^1_a \frac{3\left(1+\omega(a)\right)}{a}da\right),
\end{equation}
An SH perfect fluid is a fundamental matter field since it can be derived by a fundamental matter Lagrangian. In the following two subsections, we will describe two distinct ways of obtaining such a matter field.
\subsection{Scalar field collapse}
We prove that any SH perfect fluid is equivalent to a SH scalar field $\phi(a)$ with a suitable potential $V_s(a)$, as far as the gravitational collapse is concerned. If $\phi(a)$ is invertible, then the following statement holds: Any SH perfect fluid is gravitationally equivalent to a SH scalar field $\phi$ with a suitable potential $V_s(\phi)$. 

Consider a real scalar field defined on the manifold $\mathcal{M}$ as
\begin{equation}
    \phi: \mathcal{M} \to \mathbb{R}.
\end{equation}
The Lagrangian of a massless scalar field $\phi$ with potential $V_s(a)$ is given by
\begin{equation}
    \mathcal{L}_{\phi}=\frac{1}{2}g^{\mu \nu}\partial_{\mu}\phi \partial_{\nu} \phi-V_{s}(\phi),
\end{equation}
The stress-energy tensor is obtained from the Lagrangian $\mathcal{L}_{\phi}$ as 
\begin{equation}
    T_{\mu\nu}=-\frac{2}{\sqrt{-g}}\frac{\delta \left(\sqrt{-g} \mathcal{L}_{\phi}\right)}{\delta g^{\mu\nu}}.
\end{equation}
The density ($\rho_s$) and the isotropic pressure ($p_s$) are subsequently expressed in terms of the time derivative of the scalar field and its potential as
\begin{equation}\label{rhophi}
    \rho_s= \frac{1}{2}\dot \phi^2+V_{s}
\end{equation}
and
\begin{equation}\label{pphi}
    p_s=\frac{1}{2}\dot \phi^2-V_{s}.
\end{equation}
The overhead dot denotes the time derivative of the functions. 
From Eq.(\ref{rhophi}) and Eq.(\ref{pphi}), and from using the chain rule $\dot  \phi=\phi_{,a} \dot a$, we get
\begin{equation}\label{rhop}
    \rho_s+p_s=\phi_{,a}^2\dot a^2.
\end{equation}
We now equate $\rho_s=\rho$ and $p_s=p$. Using Eq.(\ref{pF}) and (\ref{rhop}), along with replacing $\dot a$ and $\ddot a$ using Eq.(\ref{adot}) and (\ref{ddota}), one obtains the expression of density as a function of $a$ as
\begin{equation}\label{rho(a)}
    \rho_s=\rho_0 \exp{\left(\int_a^1 a \phi_{,a}^2 da\right)}.
\end{equation}
From Eqs.(\ref{rhophi}) and (\ref{pphi}), we get
\begin{equation}\label{rhop3}
    p_s=\rho_s-2V_{s}.
\end{equation}
Using Eq.(\ref{adot}) in Eq.(\ref{rhop}), we get
\begin{equation}\label{rhop2}
    \rho_s\left(1-\frac{\phi_{,a}^2a^2}{3}\right)+p_s=0.
\end{equation}
Using Eqs.(\ref{rhop3}) and (\ref{rhop2}), we get
\begin{equation}\label{rhofinal}
    V_{s}(\phi)=\rho_s\left(1-\frac{\phi_{,a}^2a^2}{6}\right).
\end{equation}
Using Eq.(\ref{rhop}), Eq.(\ref{adot}) in Eq.(\ref{pF}), one obtains 
\begin{equation}\label{rhoabyrho}
    \frac{\rho_{s,a}}{\rho_s}=-\frac{\phi,_a^2}{a}.
 \end{equation}
We have, using Eq.(\ref{eosp}), Eq.(\ref{rhophi}) and Eq.(\ref{pphi}), 
\begin{equation}\label{vrhoomega}
    V_{s}=\frac{\rho_s}{2}\left(1-\omega \right).
\end{equation}
Now from Eq.(\ref{rhofinal}) and Eq.(\ref{vrhoomega}), we have
\begin{equation}\label{phiaomega}
    \phi(a),_{a}=\pm \frac{\sqrt{3\left(1+\omega(a)\right)}}{a}.
\end{equation}
Integrating the above equation, one obtains
\begin{equation}\label{phiomega}
    \phi(a)=\pm \int_a^1\frac{\sqrt{3\left(1+\omega(a)\right)}}{a}da+c.
\end{equation}
From Eq. (\ref{rhoomega}) and Eq.(\ref{rhofinal}) we have
\begin{equation}\label{vomega}
    V_{s}(a)=\rho_0 \left(\frac{1-\omega(a)}{2}\right)\exp\left( \int^1_a \frac{3\left(1+\omega(a)\right)}{a}da\right).
\end{equation}
\\ \\
Hence, we proved that given the functional form of the equation of state parameter $\omega(a)$, one could obtain the corresponding scalar field $\phi(a)$ given by Eq. (\ref{phiomega}) with potential $V_s(a)$ given by Eq. (\ref{vomega}). As long as $\phi(a)$ is invertible (or, in other words, a bijective map from $(0,1]\to \mathbb{R}$), we obtain $a(\phi)$, at least in principle, using which, we get $V_s(\phi)$.

Alternatively, given a scalar field $\phi(a)$, one can obtain the corresponding perfect fluid $\hat{T}$ (or the $\omega (a)$ by which it is identified), using Eq.(\ref{phiaomega}).

On the other hand, we can also start with a given scalar field potential $V(\phi)$. One can use Eq.(\ref{rho(a)}) and Eq.(\ref{rhofinal}) to obtain the ordinary nonlinear differential equation
\begin{equation}
    \mathcal{H}\left(a, \phi, \frac{d\phi}{da}, \frac{d^2\phi}{da^2}\right)=0,
\end{equation}
that can be solved in principle, to obtain $\phi(a)$, and later obtain $\omega(a)$ using Eq.(\ref{phiaomega}). Hence, given a scalar field potential $V_s(\phi)$, one can obtain the corresponding $\hat{T}$ (identified by $\omega(a)$) in the above manner. 
\\
\begin{figure*}
\begin{center}
\includegraphics[width=12.5cm]{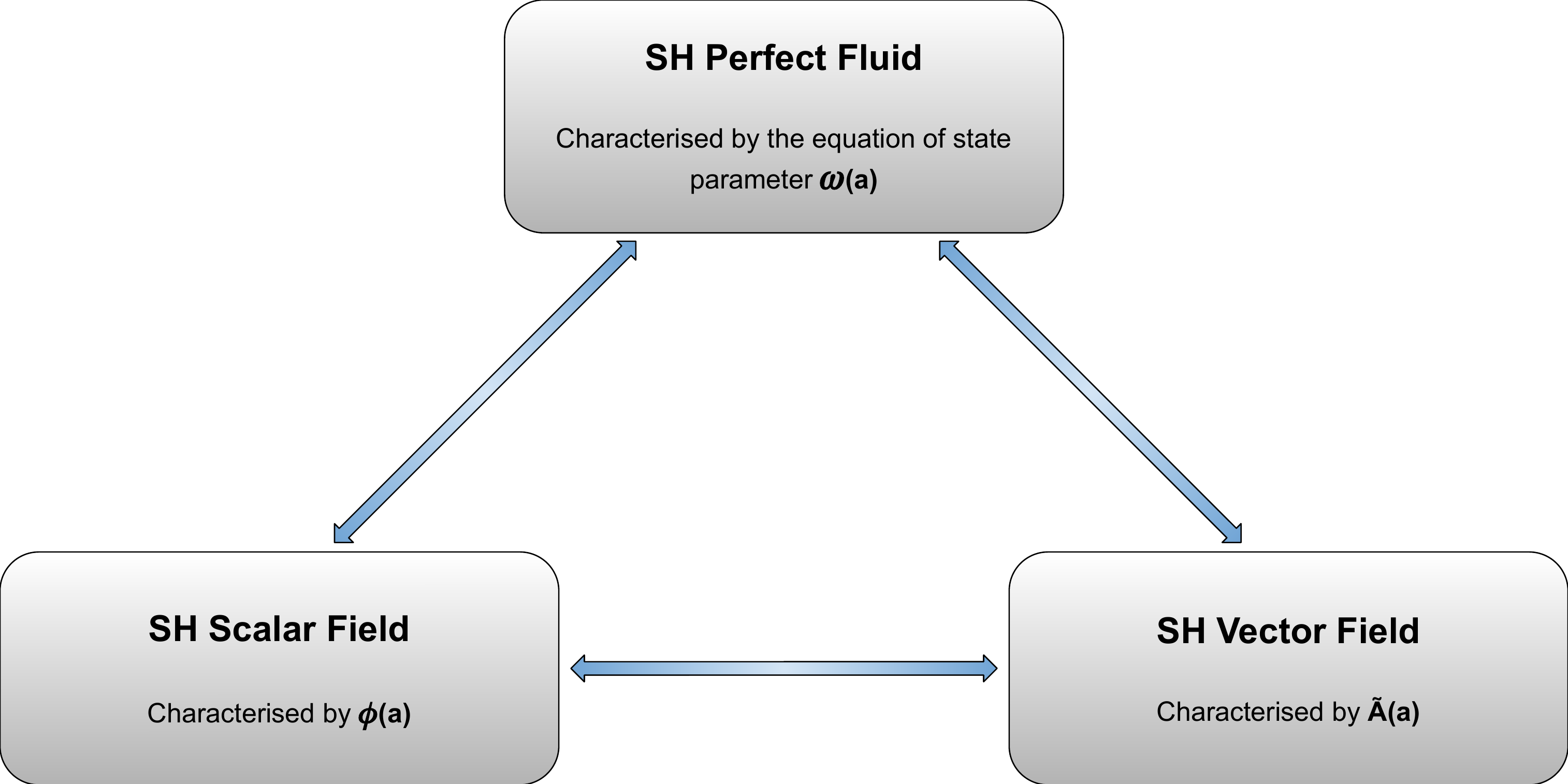}
\caption{A spatially homogeneous (SH) perfect fluid (governed by a flat FLRW spacetime metric) is completely characterized by the equation of state parameter $\omega(a)$, Eq.(\ref{eosp}) of the matter field. This matter field is obtained from fundamental matter Lagrangian. Hence, the same matter field is also characterized by an SH scalar field $\phi(a)$, Eq.(\ref{phiomega}) or its potential $V_s(a)$, Eq.(\ref{vomega}) ($V_s(\phi)$ if $\phi(a)$ is invertible). Similarly, it can also be characterized by an SH vector field  
$\Tilde{A}(a)$, Eq.(\ref{deA}) or its potential $V_v(a)$, Eq.(\ref{vg1})  ($V_v(B)$ if $B(a)$ is invertible). This schematic diagram depicts the equivalence between the gravitational collapse of SH Perfect fluid,  Scalar field and  Vector field. By spatial homogeneity, we mean homogeneous on a three-dimensional spacelike orbit with a six-dimensional isometry group $G_6$ corresponding to the spacetime \cite{Wainwright_2005}.}
\label{all}
\end{center}
\end{figure*}

\subsection{Vector field collapse}
We prove that any SH perfect fluid is equivalent to a SH vector field $\Tilde{A}(a)$ with a suitable potential $V_v(a)$, as far as the gravitational collapse is concerned. If $B(a)$ is invertible, then the following statement holds: Any SH perfect fluid is 
 gravitationally equivalent to a SH vector field $\Tilde{A}$ with a suitable potential $V_v(B)$
(where $B=g(\Tilde{A}, \Tilde{A})$). 

Consider a vector field
\begin{equation}
    \Tilde{A}: \mathcal{M} \to T\mathcal{M}.
\end{equation}
with potential $V(B)$. For a fixed $p\in \mathcal{M}$, $\Tilde{A}(p)=A_{\mu}dx^{\mu}$, where $A_{\mu}=(A_0,A_i)$, $1<i<3$ (in the comoving cartesian coordinate basis). Here $B=g^{\alpha \beta} A_{\alpha} A_{\beta}$. We consider a SH pure vector field: $A_0=0$ and $A_i=A \in \mathbb{R}$ $\forall i \in (1,2,3)$. For such a vector field, $B=3A^2/a^2$. 
 
The Lagrangian of a massless vector field $\Tilde{A}$ with potential $V_v(B)$ is given by
\begin{equation}
    \mathcal{L}_{\Tilde{A}}=-\frac{1}{4}F^{\mu \nu}F_{\mu \nu}-V_{v}(B).
\end{equation}
$F$ is a two form called the field strength and can be written in terms of wedge product as $F=F_{\mu \nu}dx^{\mu}\wedge dx^{\nu}$. The field strength is the exterior derivative of the vector field $\Tilde{A}$, i.e.$F=d\Tilde{A}$. The components are written as $F_{\mu \nu}=\nabla_{\mu}A_{\nu}-\nabla_{\nu}A_{\mu}$.

The stress-energy tensor is obtained from the Lagrangian $\mathcal{L}_{\Tilde{A}}$ as
\begin{equation}
    T_{\mu\nu}=-\frac{2}{\sqrt{-g}}\frac{\delta \left(\sqrt{-g} \mathcal{L}_{\Tilde{A}}\right)}{\delta g^{\mu\nu}}.
\end{equation}
This gives us
\begin{equation}
    T_{\mu\nu}=-\frac{1}{4}F_{\alpha \beta}F^{\alpha \beta}g_{\mu\nu}-V_{v}(B)g_{\mu\nu}+F_{\mu \alpha}F_{\nu}^{\hspace{0.15cm}\alpha}+2V_{v}'A_{\mu}A_{\nu}.
\end{equation}
The overhead prime denotes the ordinary derivative with respect to B. The density and the isotropic pressure are subsequently expressed in terms of the time derivative of the vector field component and its potential as
\begin{equation}\label{rho}
    \rho_v= \frac{3}{2}\frac{\dot A^2}{a^2}+V_{v}(B), 
\end{equation} 
and
\begin{equation}\label{p}
    p_v=\frac{1}{2}\frac{\dot A^2}{a^2} - V_{v}(B) +2V_{v}' \frac{A^2}{a^2}.
\end{equation}
\\ \\
We now equate $\rho_v=\rho$ and $p_v=p$. From Eq.(\ref{rho}) and Eq.(\ref{rhoF}), we obtain
\begin{equation}\label{vg}
    V_{v}=\rho_v\left(1-\frac{1}{2}A,_a^2\right).
\end{equation}
Substituting for $\rho(a)$ from Eq.(\ref{rhoomega}), we obtain
\begin{equation}\label{vg1}
     V_{v}=\rho_0\exp\left( \int^1_a \frac{3\left(1+\omega(a)\right)}{a}da\right)\left(1-\frac{1}{2}A,_a^2\right)
\end{equation}
On differentiating Eq.(\ref{vg}) with respect to $B$ we obtain,
\begin{equation}\label{V'}
    V_{v}'= \frac{\rho_{v,a}\left(1-\frac{A,_a^2}{2}\right) -\rho_v A,_a A,_{aa}}{\frac{6A^2}{a^2}\left(\frac{A,_a}{A}-\frac{1}{a}\right)}
\end{equation}
Using Eq.(\ref{p}), Eq.(\ref{rhoF}), and Eq.(\ref{pF}), we obtain
\begin{equation}\label{arhorhoaAAaVV'}
    \frac{a\rho_{v,a}}{3}+\rho_v\left(1+\frac{1}{6}A,_a^2\right)=V_{v}-V_{v}' \frac{A^2}{a^2}.
\end{equation}
Substituting for $V_{v}$ and $V_{v}'$ from Eq.(\ref{vg}) and Eq.(\ref{V'}), and also substituting for $\rho,_a$ (by differentiating  Eq.(\ref{rhoomega})) in Eq.(\ref{arhorhoaAAaVV'}), we obtain a second order nonlinear differential equation
\begin{equation}\label{deA}
    \mathcal{G}\left(a, \omega, A,\frac{dA}{da},\frac{d^2A}{da^2}\right)=0,
\end{equation}
where $\mathcal{G}$ is
\begin{equation}
\begin{split}
  \mathcal{G}= &\frac{d^2A}{da^2}-\frac{4}{A}\left(\frac{dA}{da}\right)^2+\frac{1}{2a}\left(5-3\omega\right)\frac{dA}{da}+\frac{6}{A}\left(1+\omega\right)\\
  &-\frac{3\left(1+\omega\right)}{a}\left(\frac{dA}{da}\right)^{-1}.    
\end{split}
\end{equation}
For a fixed $\omega(a)$, solving this differential equation with two initial conditions gives us $A(a)$, and consequently, the vector field $\Tilde{A}$. 

Hence, we proved that given the functional form of the equation of state parameter $\omega(a)$, one could obtain the corresponding vector field $\Tilde{A}$ using Eq.(\ref{deA}), and consequently, the vector field potential $V_v(a)$ using Eq.(\ref{vg1}). Now, from the functional form $A(a)$, we obtain $B(a)$. As long as $B(a)$ is invertible (or, in other words, a bijective map from $(0,1]\to \mathbb{R}$), we obtain $a(B)$, at least in principle, using which, we get $V_v(B)$.

Alternatively, given a vector field $\Tilde{A}(a)$, one can obtain the corresponding perfect fluid $\hat{T}$ (or the $\omega (a)$ by which it is identified), using Eq.(\ref{deA}).

On the other hand, we can also start with a given vector field potential $V_v(B)$. One can differentiate Eq.(\ref{vg1}), and do some rearrangements to obtain 
\begin{equation*}
\omega(a, A, \frac{dA}{da}, \frac{d^2A}{da^2})    
\end{equation*}
as
\begin{equation}\label{omegaV}
    \omega= \frac{2AV'}{a V}\left(\frac{A}{a}-\frac{dA}{da}\right)-\frac{a}{3}\frac{dA}{da}\frac{d^2A}{da^2}\left(1-\frac{1}{2}\left(\frac{dA}{da}\right)^2\right)^{-1}-1.
\end{equation}
Substituting Eq.(\ref{omegaV}) in Eq.(\ref{deA}), we obtain
\begin{equation}\label{findingA}
    \Tilde{\mathcal{G}}\left(a, A,\frac{dA}{da},\frac{d^2A}{da^2}\right)=0.
\end{equation}
In principle, this differential equation can be solved to obtain $A(a)$, which, when substituted in Eq.(\ref{omegaV}), gives us $\omega(a)$. Hence, given a vector field potential $V(B)$, one can obtain the corresponding $\hat{T}$ (identified by $\omega(a)$) in the above manner.

\section{Exterior generalized Vaidya spacetime}

The collapsing vector field spacetime ($ g_{\mu\nu}^{-}$) can be joined smoothly with the exterior generalized Vaidya spacetime ($g_{\mu\nu}^{+}$) so that their union forms a valid solution of the Einstein's field equations. The interior FLRW and the exterior generalized Vaidya spacetime 
\cite{Wang_99}
are respectively given as
\begin{equation}
    ds^2_{-}=-dt^2 + a(t)^2dr^2+r_b^2a(t)^2d\Omega^2,
\end{equation}
and
\begin{equation}
    ds^2_{+}=-\left(1-\frac{2\mathcal{M}(\mathcal{R},v)}{\mathcal{R}}\right)dv^2-2dvd\mathcal{R}+\mathcal{R}^2d\Omega^2.
\end{equation}
Here, $v$ is the retarded null coordinate, $\mathcal{R}$ is the generalized Vaidya radius, and $r_b$ is the value of the radial coordinate $r$ corresponding to the matching hypersurface, or in other words, the radial coordinate of the outermost shell of the collapsing scalar/vector field cloud. The matter field corresponding to the generalized Vaidya spacetime is a combination of \textit{Type I} and \textit{type II}, such that the components of the stress-energy tensor written in the orthonormal basis appear as
\begin{equation}
T_{a b}=
    \begin{pmatrix}
\frac{\Bar{\epsilon}}{2}+\epsilon & \frac{\Bar{\epsilon}}{2} & 0 & 0\\
\frac{\Bar{\epsilon}}{2} & \frac{\Bar{\epsilon}}{2}-\epsilon & 0 & 0 \\
0 & 0 & \mathcal{P} & 0\\
0 & 0 & 0 & \mathcal{P}.
\end{pmatrix}
\end{equation}
$\epsilon=\mathcal{P}=0$ and $\Bar{\epsilon}\neq 0$ corresponds the usual Vaidya spacetime as a special case. $\Bar{\epsilon}=0$ and $\epsilon \neq 0$ corresponds to a sub-class of \textit{Type I} matter field. The generalized Vaidya solution encompasses many known Einstein field equations solutions.
Matching the first and second fundamental forms for the interior and exterior metric on $\Sigma$ gives the following equations:
\begin{equation}\label{mc1}
    \mathcal{R}(t)=R(t,r_b) \left(=r_b a(t)\right),
\end{equation}
\begin{equation}\label{mc2}
    F(t,r_b)=2\mathcal{M}(\mathcal{R},v),
\end{equation}
\begin{equation}\label{mc3}
    \left(\frac{dv}{dt}\right)_{\Sigma}=\frac{1+\dot {\mathcal{R}} }{1-\frac{F(t,r_b)}{\mathcal{R}}},
\end{equation}
and
\begin{equation}\label{mc4}
    \mathcal{M}(\mathcal{R},v),_{\mathcal{R}}=\frac{F(t,r_b)}{2 \mathcal{R}}+\mathcal{R} \ddot {\mathcal{R}}.
\end{equation}
Here, $F=R\dot R^2$ is the Misner-Sharp mass function of the collapsing spherical SH perfect fluid. Using the relation (\ref{mc1}), we can relate the generalized Vaidya mass with the density of the interior collapsing SH spherical perfect fluid cloud as
\begin{equation}\label{mrho}
    \mathcal{M}=\frac{\rho}{6}\mathcal{R}^3.
\end{equation}
Using Eq.(\ref{ddota}), differentiation of Eq.(\ref{phiomega}) with respect to $a$, and Eq.(\ref{mrho}), in Eq.(\ref{mc4}) we get
\begin{equation}\label{momega}
  \mathcal{M},_\mathcal{R} = \frac{3\mathcal{M}}{\mathcal{R}} \left( 1+\frac{(1+\omega (a))r^2_b}{\mathcal{R}^2}\right),
\end{equation}
integrating which we obtain
\begin{equation}\label{GV1}
    \mathcal{M}(\mathcal{R},v)=\mathcal{M}_1 (v) \exp \left(\int \frac{3 }{\mathcal{R}}\left( 1+\frac{(1+\Tilde{\omega} \left(\mathcal{R}\right))r_b^2}{\mathcal{R}^2}\right) d\mathcal{R}\right).
\end{equation}
Here $\mathcal{M}_1(v)$ is a constant of integration and is a function of null coordinate $v$, and 
\begin{equation*}
\Tilde{\omega}(\mathcal{R})=\omega\left(\frac{\mathcal{R}}{r_b}\right).    
\end{equation*}
Eq.(\ref{GV1}) gives us the expression of the generalized Vaidya mass function of the exterior generalized Vaidya spacetime, in terms of interior collapsing perfect fluid equation of state parameter $\omega$, to ensure smooth matching at the matching hypersurface. 

For the exterior matter field to satisfy the weak energy condition, $\bar {\epsilon}$ and $\epsilon$ should be non-negative
\cite{Wang_99}. 
These inequalities, in turn, put restrictions on the generalized Vaidya mass function as
\begin{equation}
    \mathcal{M},_v \leq 0, \hspace{0.5cm} \textrm{and} \hspace{0.5cm}  \mathcal{M},_{\mathcal{R}} \geq 0.
\end{equation}
Using Eq.(\ref{momega}) and Eq.(\ref{GV1}) in the above two relations, we obtain
\begin{equation}\label{ec1}
    \mathcal{M}_{1,v}\leq 0,
\end{equation}
and
\begin{equation}\label{ec2}
\left(\frac{1+\omega(a)}{a^2}\right)\geq 0.
\end{equation} 
The inequality (\ref{ec2}) is always satisfied if the interior collapsing matter field obeys the weak energy condition. Hence, Eq.(\ref{ec1}) is the only restriction on the generalized Vaidya mass function for the exterior spacetime to obey at least the weak energy condition. 

Now, we have a complete solution of Einstein's field equations consisting of an interior collapsing SH scalar/vector field (with some potential) and the exterior generalized Vaidya solution, matched smoothly at the matching hypersurface. The free functions are categorically the potential function ($V_s(\phi)$ in case of scalar field collapse, and $V_{v}(B)$ in case of vector field collapse), and the component of generalized Vaidya mass function $\mathcal{M}_1(v)$, the latter one restricted by the inequality (\ref{ec1}). It is evident that the choice of $\mathcal{M}_1(v)$ does not affect the causal structure of the spacetime obtained as an end-state of unhindered gravitational collapse. Of course, instead of considering the potential function $V_s(\phi)$ (or $V_v(B)$) as a free function, one could also consider any one of the remaining functions: $\omega(a)$, $\rho(a)$, $\phi(a)$ (or $A(a)$), $V_{s}(a)$ (or $V_v(a)$) as a free function, without any trouble. In the next section, we study the end state of this class of global dynamical spacetime identified by any one of the free functions.

\section{Causal Structure and strength of the singularity}

Once the singularity is formed as an end state of gravitational collapse of the interior scalar (vector) field with potential $V_s(\phi)$ ($V_v(B)$), one can investigate \textcolor{black}{whether or not causal geodesics can escape the singularity. Additionally, one can investigate whether or not such singularity is gravitationally strong in the sense of Tipler. The following two subsections discuss these two properties.} 

    \begin{table*}
        \begin{center}
\begin{tabularx}{1\textwidth} {|p{62mm}|p{36mm}|p{40mm}|p{16.5mm}|p{10mm}|}
 \hline
  Massless scalar field & $V_s(\phi)=0$  & $\phi(a)=c\pm \sqrt{6} \log a $  & strong & BH \\ 
 \hline
  Homogeneous dust ($\omega=0$) & $V_s(\phi)\propto \exp \left(\sqrt{3}\phi\right)$ & $\phi(a)=c\pm \sqrt{3}\log a$ & strong & BH \\ 
 \hline
 Goswami/ Joshi \cite{goswami2007} ($\omega=-\frac{2}{3}$) (SF1) & $V_s(\phi) \propto \exp \phi$ & $\phi(a)=c \pm \log a$ & strong & NS\\
 \hline
 Two dimensional analog of Mexican hat \cite{Mosani2022} (SF2) & $V_s(\phi)=\frac{1}{2}\mu \phi^2+\lambda \phi^4$ & $\phi(a)=\pm 2\sqrt{2}\sqrt{c-\log a}$ & weak & NS\\
 \hline
\end{tabularx}
\end{center}
\caption{Four examples of spatially homogeneous scalar fields that collapse to form a singularity that is either hidden (blackhole or BH) or (naked singularity or NS). In the fourth example, $\mu=-\frac{16}{3} \lambda$. The first three types end up in a strong singularity in the sense of Tipler.}
\label{table:1}
\end{table*}
\begin{table*}
  \begin{center}
\begin{tabularx}{1\textwidth} {|p{101mm}|p{38mm}|p{18.5mm}|p{10mm}|}
 \hline
  Massless vector field & $V_v(B) =0$ & strong & BH \\ 
 \hline
 Massive vector field & $V_v(B) =-\frac{1}{2}\mu^2 B$ & strong & BH\\ 
 \hline
 \textcolor{black}{VF1} & $V_v(a)$ as in Fig.(3) & strong & NS\\
 \hline
  \textcolor{black}{VF2} & $V_v(a)$ as in Fig.(3)  & weak & NS \\
 \hline
\end{tabularx}
\end{center}
\caption{Four examples of spatially homogeneous vector fields that collapse to form a singularity that is either hidden within a black hole (BH) or is naked (NS). The ones mentioned in the third and the fourth row are newly constructed vector fields from known scalar fields (mentioned in the third \cite{goswami2007} and the fourth \cite{Mosani2022} row of Table 1, respectively) by exploiting the gravitational equivalence depicted in Fig.(1). The corresponding vector field component $A(a)$ for each case is plotted in Fig.(2-3). The first three types end up in a gravitationally strong singularity in the sense of Tipler.}
\label{table:2}
\end{table*}
\begin{center}
\begin{figure*}
    \begin{subfigure}[b]{0.47\textwidth}
        \includegraphics[width=\textwidth]{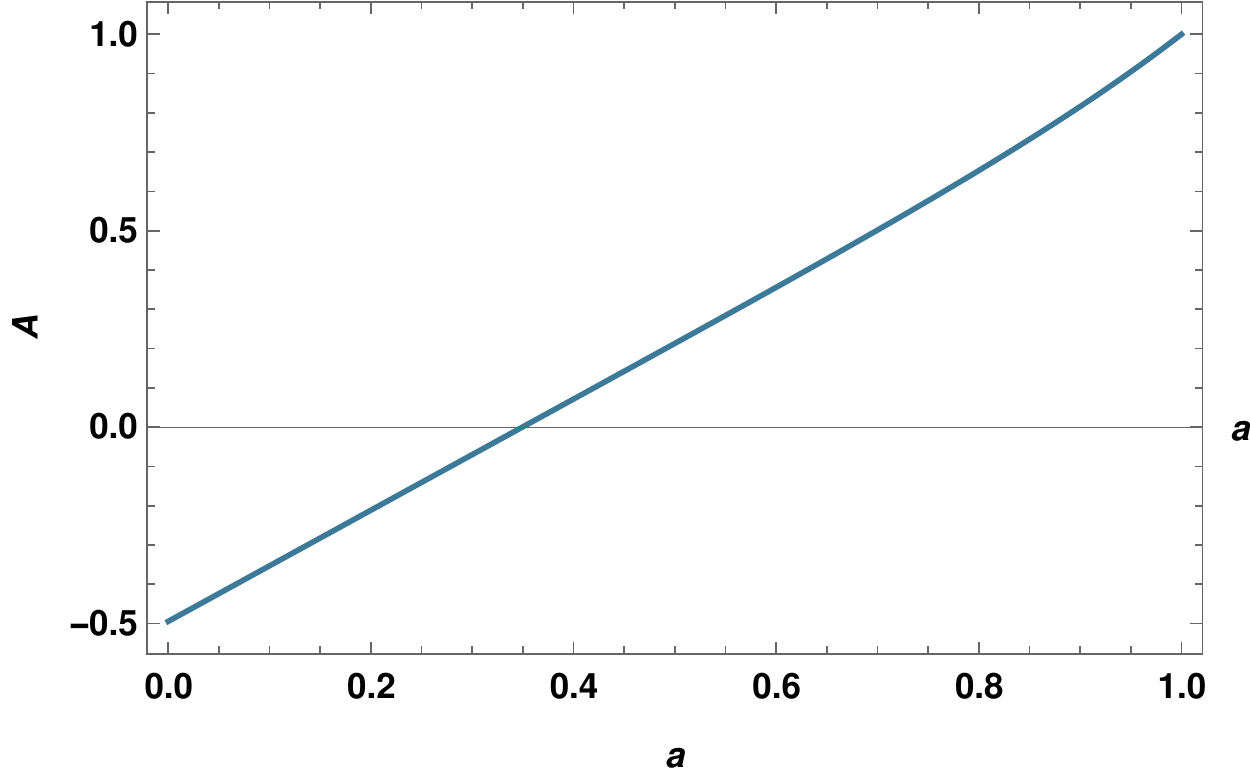}
        \label{rfidtest_xaxis}
    \end{subfigure}
    \begin{subfigure}[b]{0.44\textwidth}
        \includegraphics[width=\textwidth]{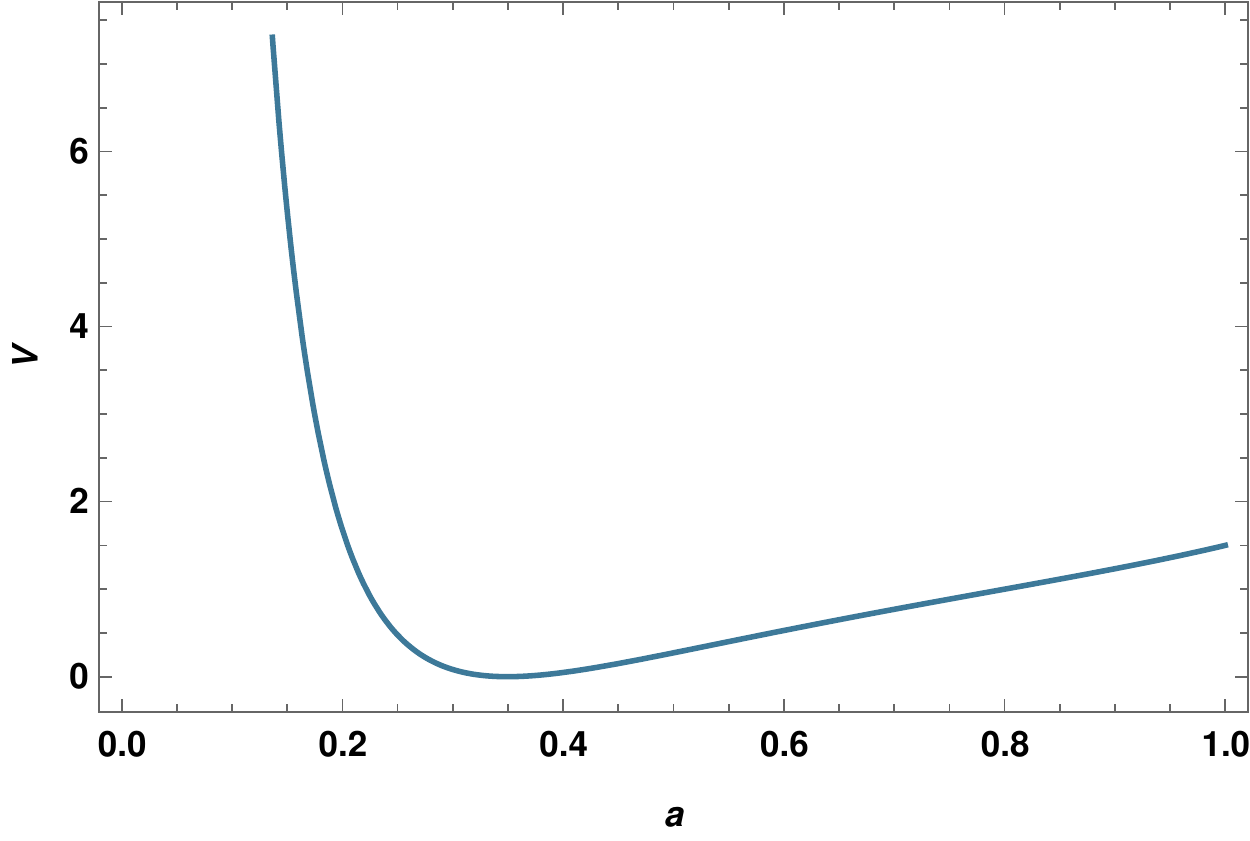}
        \label{rfidtest_yaxis}
    \end{subfigure}
    \caption{\textcolor{black}{The dynamics of the vector field component $A(a)$ in the case of the massive ($\mu=1$) vector field $\Tilde{A}$ (Left panel) and its potential $V_{v}(a)$ (Right panel). First we obtain $\omega(a, A, \frac{dA}{da}, \frac{d^2A}{da^2})$ by substituting $V_v(B)=-\frac{1}{2}\mu^2B$ in Eq.(\ref{omegaV}). Substituting for $\omega(a, A, \frac{dA}{da}, \frac{d^2A}{da^2})$ in Eq.(\ref{findingA}) and solving the differential equation with initial conditions $A(1)=1$ and $A'(1)=2$, we obtain $A(a)$. Consequently,  substituting $V_v(B)=-\frac{1}{2}\mu^2B$, and the obtained $A(a)$ in Eq.(\ref{omegaV}), we obtain $\omega(a)$, Further substitution of $\omega(a)$ in Eq.(\ref{vg1}), we obtain $V_v(a)$.}}
    \label{rfidtag_testing}
\end{figure*}
\begin{figure*}
    \centering
    
    \begin{subfigure}[b]{0.45\textwidth}
        \includegraphics[width=\textwidth]{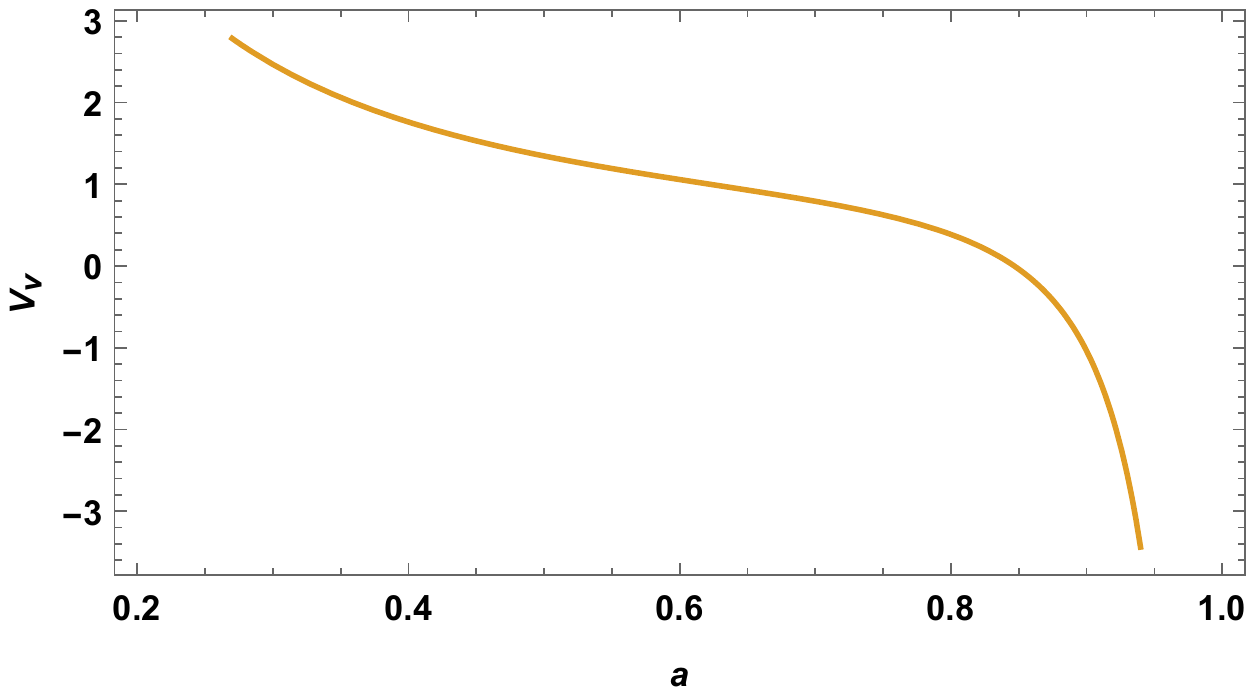}
        \caption{}
        \label{rfidtest_yaxis}
    \end{subfigure}
    \begin{subfigure}[b]{0.45\textwidth}
        \includegraphics[width=\textwidth]{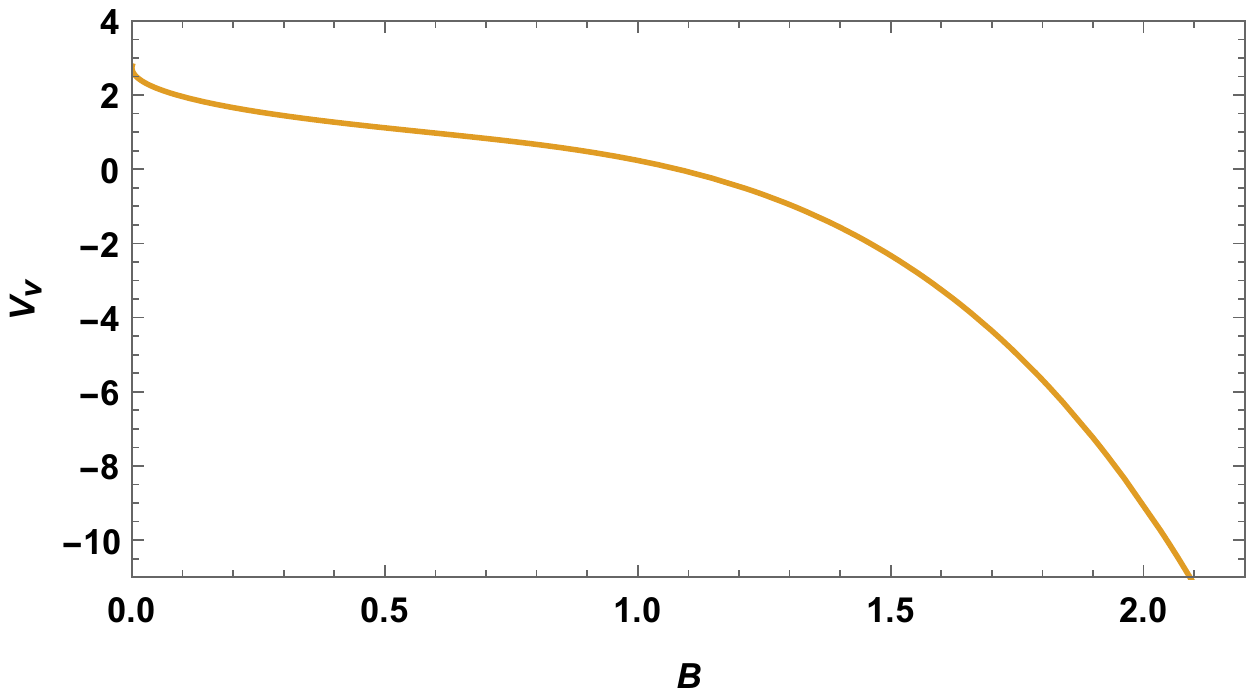}
        \caption{}
        \label{rfidtest_yaxis}
    \end{subfigure}
    \begin{subfigure}[b]{0.45\textwidth}
        \includegraphics[width=\textwidth]{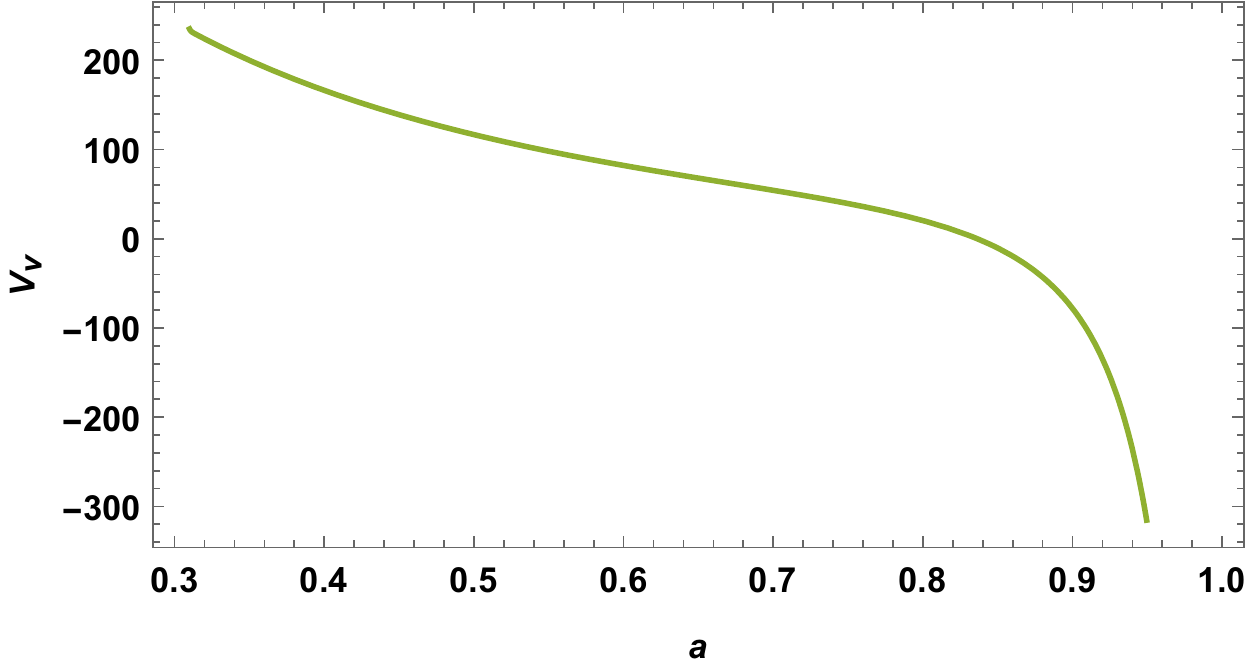}
        \caption{}
        \label{rfidtest_yaxis}
    \end{subfigure}
    \begin{subfigure}[b]{0.45\textwidth}
        \includegraphics[width=\textwidth]{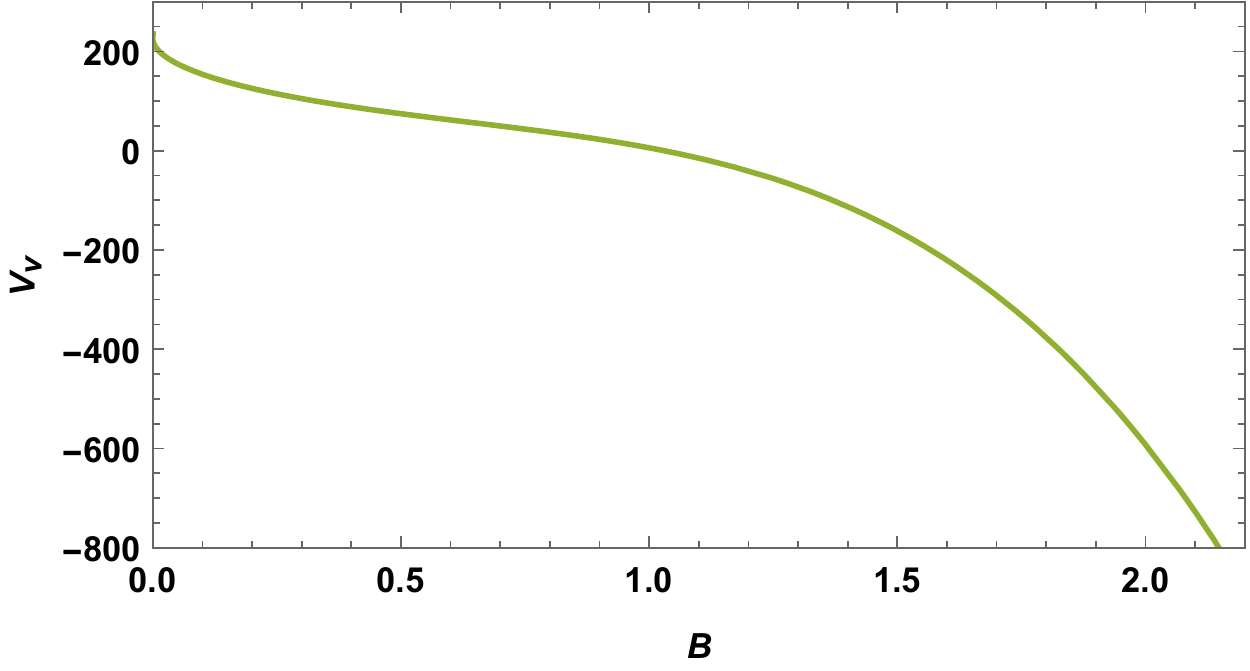}
        \caption{}
        \label{rfidtest_yaxis}
    \end{subfigure}
    \begin{subfigure}[b]{0.45\textwidth}
        \includegraphics[width=\textwidth]{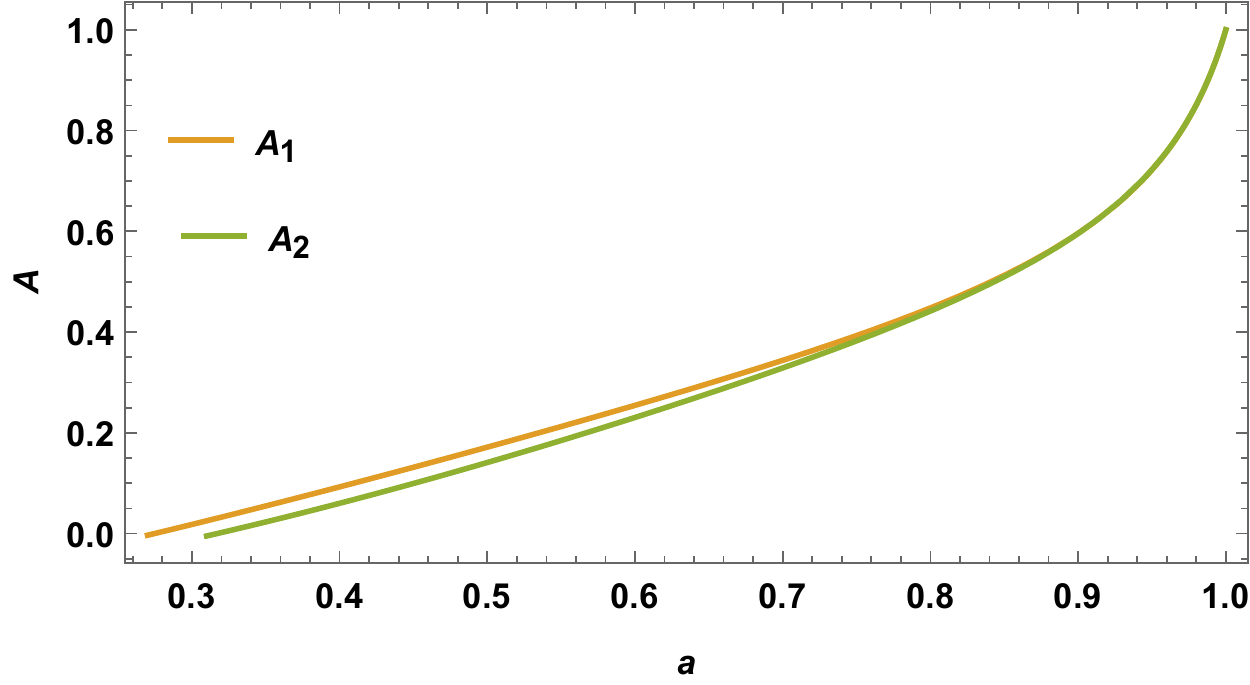}
        \caption{}
        \label{rfidtest_xaxis}
    \end{subfigure}
    \caption{ \textbf{(a)} and \textbf{(c)}: Vector field potentials $V_v(a)$ corresponding to newly constructed vector fields VF1 (orange) and VF2 (green), as mentioned in the third and fourth row of the Table (\ref{table:2}), respectively. \textbf{(b)} and \textbf{(d)}: The same vector field potentials $V_{v}(B)$ as function of $B$. \textbf{(e)}: The vector field components $A(a)$ in both of these cases. In the latter example, $\mu=-8/3$ and $\lambda=1$. 
    First, we obtain $\omega_i(a)$ , using Eq.(\ref{phiomega}) (Here $i \in {1, 2}$ corresponds to VF1 and VF2 respectively). Then we obtain the vector field components $A_i(a)$ by solving the differential Eq.(\ref{deA}) with initial conditions $A_i(1) = 1$ and $A'_i(1) = 10$. Further substitution of $\omega_i(a)$ and the obtained $A_i(a)$ in Eq.(\ref{vg1}), we get $V_{v(i)}(a)$. Once $V_{v(i)}(a)$ is obtained, we obtain $V_{v(i)}(B)$.  
    }
    \label{rfidtag_testing}
\end{figure*}
\end{center}
\begin{center}
\begin{figure*}
    \begin{subfigure}[]{0.47\textwidth}
        \includegraphics[width=\textwidth]{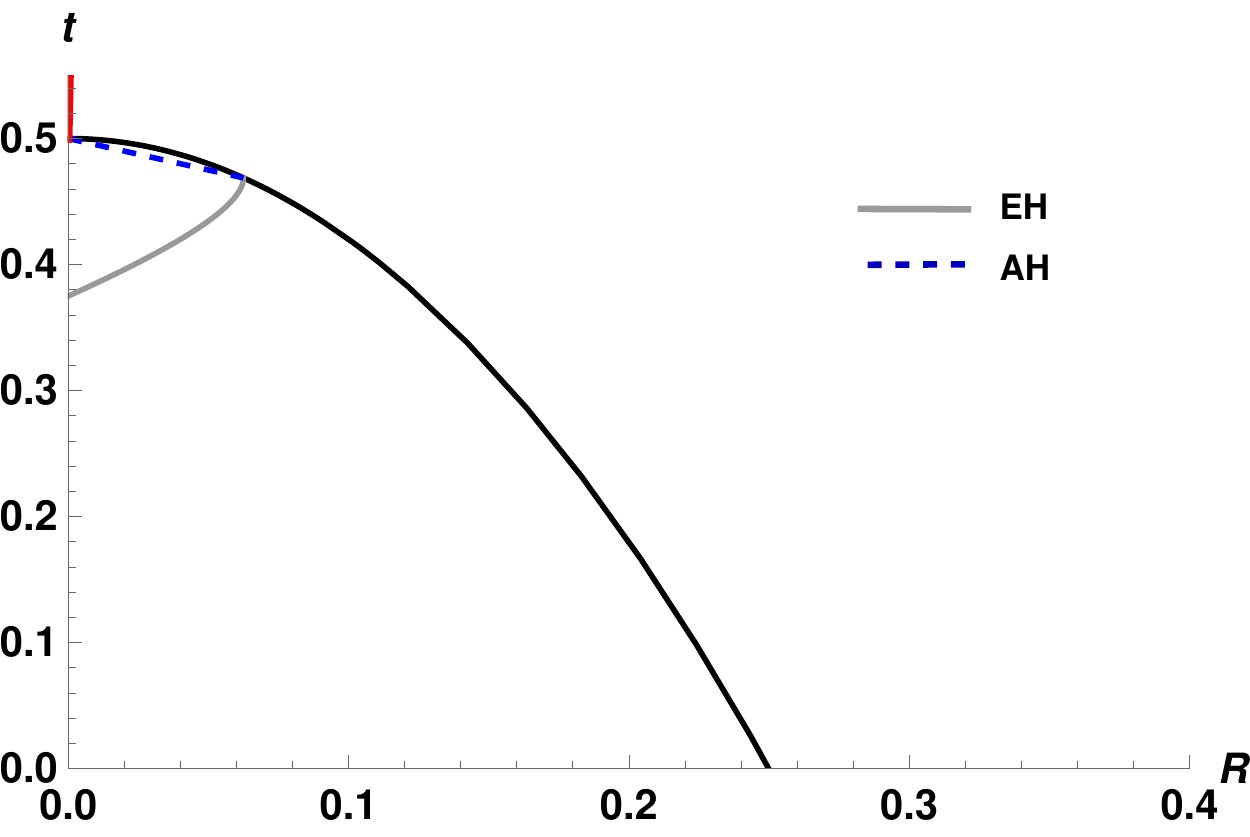}
        \label{w1}
        \subcaption{Massless scalar field ($V_s = 0$)}
    \end{subfigure}
    \begin{subfigure}[]{0.45\textwidth}
        \includegraphics[width=\textwidth]{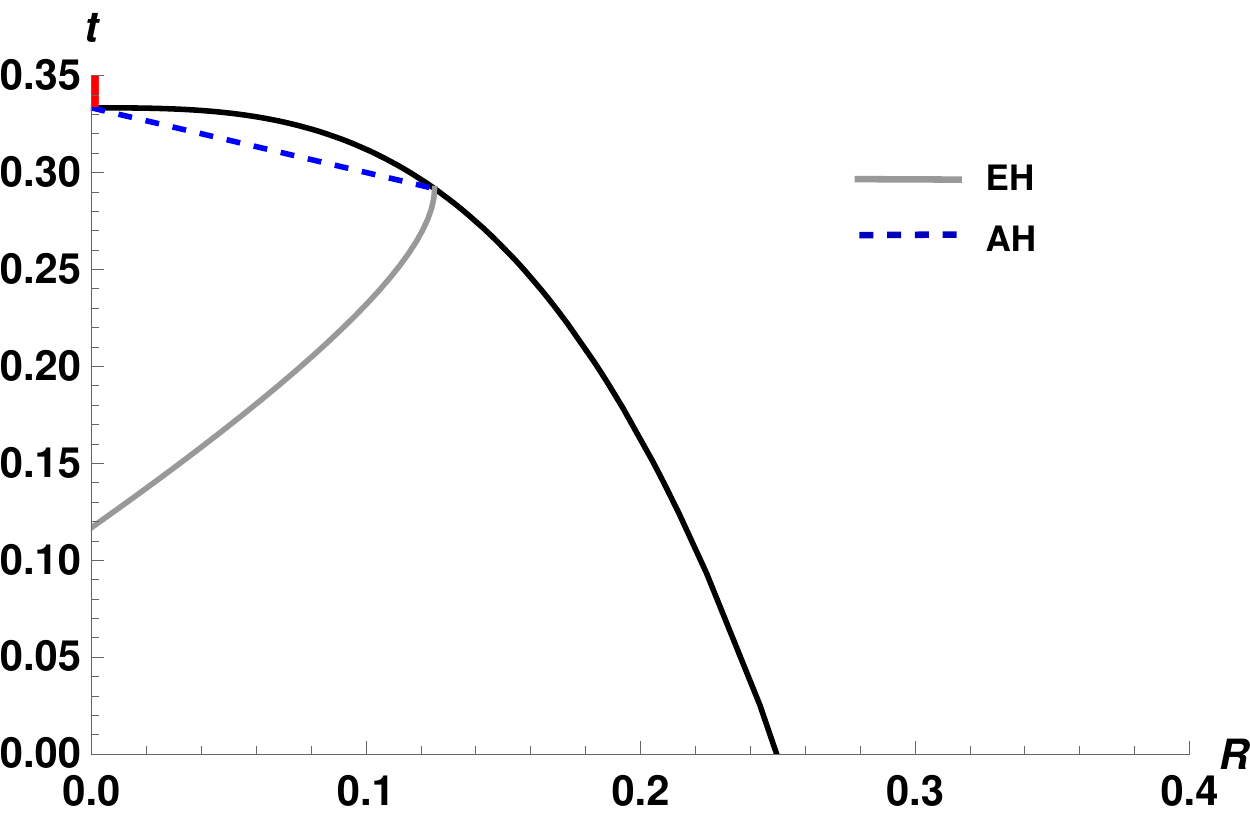}
        \label{w13}
        \subcaption{Massless vector field ($V_v = 0$)}
    \end{subfigure}
    \begin{subfigure}[]{0.45\textwidth}
        \includegraphics[width=\textwidth]{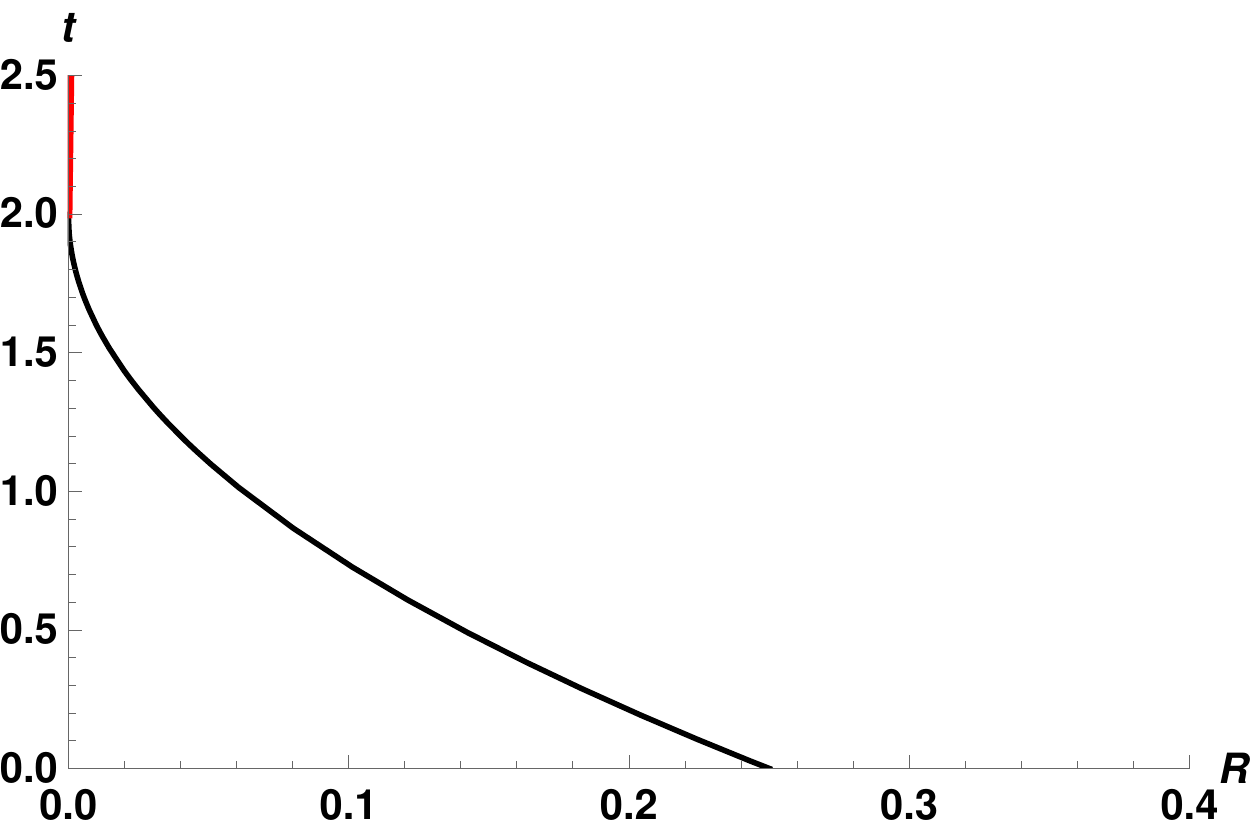}
        \label{w-23}
        \subcaption{SF1/VF1}
    \end{subfigure}
    \begin{subfigure}[]{0.45\textwidth}
        \includegraphics[width=\textwidth]{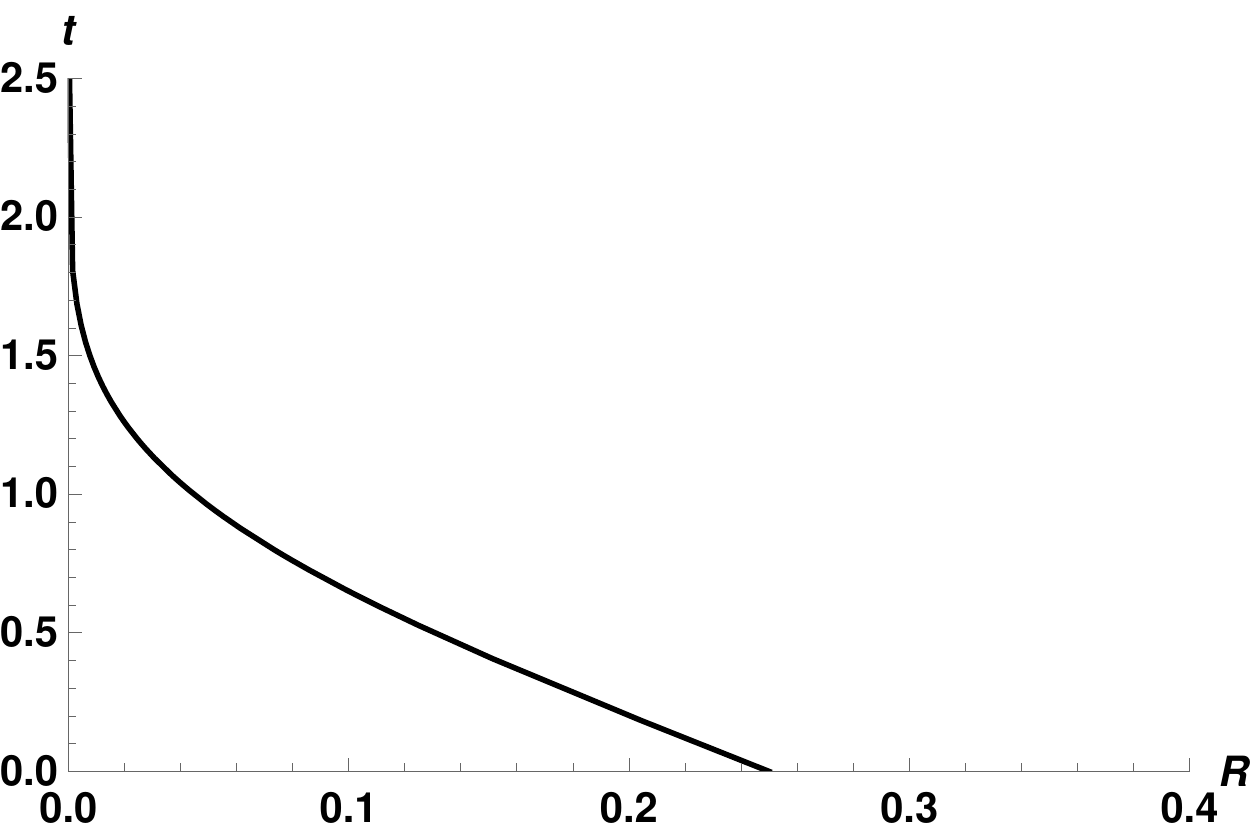}
        \label{wa}
        \subcaption{SF2/VF2}
    \end{subfigure}
    \caption{\textcolor{black}{Spacetime diagram of the examples of spatially homogeneous scalar fields and vector fields mentioned in Tables I and II. The solid black curve in each of them represents the boundary of the collapsing cloud. Upper panel: The singularity is not visible in both examples. Lower Panel: In the case of SF1/VF1, the singularity forms in a finite comoving time and is globally visible because of the absence of the apparent and event horizons. In the case of SF2/VF2, the singularity forms in an infinite comoving time. However, an ultra-high density region is obtained in finite comoving time, which can be visible globally because of the absence of the apparent and event horizons.} }
    \label{rfidtag_testing}
\end{figure*}
\end{center}
\subsection{Causal structure of the singularity}
\textcolor{black}{We say that a singularity formed due to unhindered gravitational collapse is naked if there exists a family of outgoing causal curves whose past endpoint is the singularity. In the future, these curves can either reach a faraway observer or fall back to the singularity. The singularities are then termed globally naked and locally naked, respectively. Whether or not the singularity is naked essentially depends on the geometry of trapped surfaces as the collapse evolves. Trapped surfaces are two-surfaces in the spacetime on which not only the ingoing congruence but also the outgoing congruence necessarily converge. Convergence or otherwise of the outgoing null geodesic congruence is determined by the behaviour of its expansion scalar, which we denote here as $\theta_l\left(t,r\right)$. It is expressed in terms of the metric coefficients, in comoving spherical coordinates as,
\begin{equation}
    \theta_l= \frac{2}{R}\left(1-\sqrt{\frac{\rho R^2}{3}}\right).
\end{equation}
The region in which $\theta_l<0$ is called the trapped region. The boundary of the trapped region, given by $\theta_l=0$, is called the apparent horizon. If the neighbourhood of the singular center is surrounded by a trapped region since before the time of formation of the singularity $t_s$, then it is covered, and we get a black hole. Hence, the necessary condition for singular null geodesic congruence to escape the singularity is the absence of a trapped region, which is ensured by the condition $\theta_l(t_s,r) >0$ for such congruence. 
The absence of trapped region in the neighbourhood of the singularity $(t,r)=(t_s,0)$ is ensured by the following inequality:}

\begin{equation}\label{trappedsurface}
    \lim_{t\to t_s} \frac{\rho R^2}{3} \leq \lim_{a\to 0} \frac{\rho (a) r_b^2 a^2}{3}<1,
\end{equation}
The inequality (\ref{trappedsurface}) is definitely satisfied if 
\begin{equation}\label{nts}
    \lim_{a\to 0}\rho(a)<\frac{1}{a^2}.
\end{equation}
For 
\begin{equation*}
    \lim_{a\to 0}\rho(a)=\frac{k}{a^{2}},
\end{equation*}
for some $k\in \mathbb{R}^{+}$, the inequality is satisfied only for $r_b< \sqrt{3/k}$. Rewriting the inequality (\ref{nts}) in terms of the equation of state parameter $\omega (a)$ using Eq.(\ref{rhoomega}), one obtains
\begin{equation}\label{restw}
    \lim_{a \to 0} \rho_0 a^2 \exp\left(\int^1_a \frac{3\left(1+\omega(a)\right)}{a}\right)da < 1.
\end{equation}
If a collapsing matter field with the equation of state parameter $\omega (a)$ satisfies the inequality (\ref{restw}), then it will up in a naked singularity
\cite{goswami2007}. 
In the case of otherwise, the final outcome is a black hole. 

Hence, as decided by the above inequality, we get a class of SH matter fields that include scalar and vector fields, identified by the functional form $\omega(a)$, that goes to either the blackhole or naked singularity final state as an end state of unhindered gravitational collapse.

In the case of scalar field collapse, the restriction (\ref{restw}) on $\omega(a)$ gives us a restriction on the scalar field $\phi(a)$ using Eq.(\ref{phiomega}), and the scalar field potential function $V_s(a)$ using Eq.(\ref{vomega}). Hence, obtaining a class of $\omega(a)$ is \textcolor{black}{gravitationally} equivalent to obtaining a class of scalar field potentials $V_s(a)$ that goes to the naked singularity as an end state of unhindered gravitational collapse. Moreover, suppose $\phi(a)$ is a bijective map from $(0,1] \to \mathbb{R}$. In that case, obtaining a class of $\omega(a)$ is \textcolor{black}{gravitationally} equivalent to obtaining a class of scalar field potentials $V_{s}(\phi)=V_s(a(\phi))$ that goes to the naked singularity as an end state of gravitational collapse.

Similarly, in the case of vector field collapse, the restriction (\ref{restw}) on $\omega(a)$ gives us a restriction on the vector field $\Tilde{A}$ (or more specifically, a restriction on the vector field component $A(a)$) obtained by solving the differential Eq.(\ref{deA}), and the vector field potential function $V_{v}(a)$ obtained by substituting $A(a)$ and $\rho$ from Eq.(\ref{rhoomega}), in Eq.(\ref{vg}). Hence, obtaining a class of $\omega(a)$ is \textcolor{black}{gravitationally} equivalent to obtaining a class of vector field potential $V_{v}(a)$ that goes to the naked singularity as an end state of unhindered gravitational collapse. Moreover, suppose $A(a)$ is a bijective map from $(0,1] \to \mathbb{R}$. In that case, obtaining a class of $\omega(a)$ is \textcolor{black}{gravitationally} equivalent to obtaining a class of vector field potential $V_{v}(A)=V_v(a(A))$ that goes to the naked singularity as an end state of unhindered gravitational collapse. 

\textcolor{black}{In Table ($I$) and ($II$), we discuss examples of such scalar field collapse and vector field collapse that ends up in either a black hole or a naked singularity. Exploiting the equivalence between SH perfect fluids, scalar fields with potential $V_s(a)$, and vector fields with potential $V_v(a)$, we construct two examples of collapsing vector fields with potential out-of-known examples of collapsing scalar fields with potentials, giving rise to the naked singularity as an end state.} 

\textcolor{black}{The first example of a collapsing vector field with potential $V_v(a)$ is constructed from the scalar field with potential mentioned in the third row of Table ($I$) \cite{goswami2007}. The perfect fluid corresponding to such scalar field example has an equation of state parameter $\omega(a)=-\frac{2}{3}$. The constructed collapsing vector field $\Tilde{A}=(0, A, A, A)$ (in the comoving coordinate basis) has the property (dynamics of $A(a)$ and $V_v(a)$) as shown in Fig.($3$). Refer to the third row of Table ($II$).}

\textcolor{black}{The second example of a collapsing vector field with potential $V_v(a)$ is constructed from the scalar field with potential mentioned in the fourth row of Table ($I$) \cite{Mosani2022}. Such a scalar field has a two-dimensional analogue of Mexican hat-shaped potential. The constructed collapsing vector field $\Tilde{A}=(0, A, A, A)$ (in the comoving coordinate basis) has the property (dynamics of $A(a)$ and $V_v(a)$) as shown in Fig.($3$). Refer to the fourth row of Table ($II$). The spacetime diagrams of some of the examples in Table ($I$) and ($II$) are plotted in Fig.(4).}


%
\subsection{Strength of the singularity }
Generally, a singularity in the spacetime manifold is identified by the existence of at least one past/future incomplete geodesic. However, in the case of singularities forming as the end state of a gravitational collapse, apart from the existence of such incomplete geodesics, one expects an additional physical property as follows: An object approaching such singularity should be crushed to zero volume. We call such a singularity \textcolor{black}{gravitationally \textit{strong}} in the sense of Tipler \cite{Tipler}. A precise definition of a strong singularity is as follows:

Consider a smooth spacetime manifold $(\mathcal{M},g)$ and a causal geodesic $\gamma: [t_0,0) \rightarrow \mathcal{M}$. Let $\lambda$ be an affine parameter along this geodesic. Let $\xi_{\left(i\right)}$, ($0 \leq i \leq 2$ in the case of null geodesic, $0 \leq i \leq 3$ in the case of timelike geodesic) be the independent Jacobi vector fields.
The wedge product of these Jacobi fields gives us the volume form $\mathcal{V}=\bigwedge \xi_{\left(i\right)}$. We say that a singularity is \textcolor{black}{gravitationally strong} in the sense of Tipler if this volume form vanishes as $\lambda \to 0$. 

Clarke and Krolak \cite{Klorak} 
related the existence of a \textit{Tipler} strong singularity with the growth rate of the curvature terms as follows: At least along one null geodesic with affine parameter $\lambda$ (such that $\lambda \to 0$ as the singularity is approached), the following inequality
\begin{equation}\label{strength}
\lim_{\lambda \to 0} \lambda^2 R_{ij} K^i K^j > 0
\end{equation}
should hold for the singularity to be strong in the sense of Tipler. Here $K^i=\frac{dx^i}{d\lambda}$ are the tangents to the chosen null geodesic, and $x^i$ is the coordinate system. This condition puts a lower bound on the growth of the curvature scalar. In the spherical coordinate system $(t,r,\theta,\phi)$, the radial null geodesic equation reads
\begin{equation}
    \frac{dt}{dr}=a.
\end{equation}
Hence, we have the relation between the tangents $K^t$ and $K^r$ as
\begin{equation}
    K^t=aK^r,
\end{equation}
and subsequently, in terms of the affine parameter, 
\begin{equation}
    K^t= \frac{R}{\lambda}, \hspace{0.5cm} \textrm{and} \hspace{0.5cm} K^r=\frac{r}{\lambda}.
\end{equation}
The inequality (\ref{strength}) can then be written in terms of $\omega$ as
\begin{equation}\label{strength_FLRW}
    \lim_{a \to 0} \left(r^2 (1+\omega) \rho_0 \exp\left({\int_a^1 \frac{3 (1+\omega)}{a}da}\right) \right)>0
\end{equation}
Hence, the singularity formed due to the gravitational collapse of a scalar/vector field is strong in the sense of Tipler if the following inequality holds (assuming that the weak energy condition is respected):
\begin{equation}\label{strength_omega}
    \lim_{a \to 0} \exp\left({\int_a^1 \frac{3 (1+\omega)}{a}da}\right) > 0.
\end{equation}
Hence, (along with using the condition (\ref{restw})) one can obtain a naked singularity that is strong in the sense of Tipler for that $ \omega$ that satisfies the following constraint:
\begin{equation}\label{strongnaked}
    0< \lim_{a \to 0} \exp\left({\int_a^1 \frac{3 (1+\omega)}{a}da}\right)< \mathcal{O}(a^{-2}).
\end{equation}

This constraint gives us the class of SH collapsing matter fields that we identify by $\omega(a)$, which ends up in strong curvature naked singularity. Or in other words, we have a class of scalar/vector field potentials corresponding to the given scalar/vector field that collapses to a strong naked singularity. As an example, in Tables (I) and (II), we mention the causal property and the strength of the singularity formed due to the gravitational collapse of four different scalar/vector fields. 



\section{Conclusions and Remarks}
Following are the concluding remarks:
\begin{enumerate}
    \item Unlike the singularity theorems that provide rigorous proof of the existence of incomplete causal geodesics under rather generic conditions, one does not currently have proof or disproof of the cosmic censorship hypothesis. In fact, we need a mathematically rigorous formulation of this conjecture, which is not available currently, before we can prove or disprove it. 
    
    Under the situation at present, we can only speculate its validity or otherwise. Proposed counterexamples, hence have great importance in understanding whether naked singularities, in fact, exist or not in our 
    universe. Through such analysis of gravitational collapse models only, one could possibly hope to arrive at a suitable formulation of cosmic censorship. The collapse of inhomogeneous dust and the Vaidya null fluids were the first examples proposed to produce naked singularities. However, an important objection could be that, even if astrophysically interesting, they are not fundamental forms of matter
    \cite{Eardley_1976, Joshi_Global}. 
    One could then ask whether the collapse of matter configuration that is obtained from a fundamental matter Lagrangian ends up in a naked singularity. Scalar fields with potential and vector fields with potential are fundamental matter fields in this sense. Here we show that not just one particular choice of these fields but an entire class of such types could collapse and form a naked singularity as an end state. This basically divides the allowed class of potential functions
    into classes that take the unhindered collapse to a black hole or naked singularity.  

    \item\textcolor{black}{To achieve this, we show equivalence between SH
    \begin{enumerate}
        \item \textit{Perfect fluid}: characterized by $\omega(a)$,
        \item \textit{Massless scalar field $\phi$}: characterized by $\phi(a)$ or its potential $V_s(a)$ or $V_s(\phi)$ (if $\phi(a)$ is invertible), and
        \item \textit{Massless vector field $\Tilde{A}$}: characterized by $A(a)$, or its potential $V_v(a)$, or $V_v(B)$ (if $B(a)$ is invertible). 
    \end{enumerate}
    as far as the gravitational collapse is concerned. This gravitational equivalence is described in subsections of section ($II$) and depicted in Fig.($1$). Now, if the functional form of $\omega(a)$  satisfies the inequality (\ref{restw}), then the singular null geodesic congruence, if at all there exists, does not get trapped as $a\to 0$. Hence, we have a class of functions $\omega(a)$ corresponding to a naked singularity as an end state of gravitational collapse. Now, because of the above equivalence, in the case of an SH scalar field collapse, one then has a class of scalar field function $\phi(a)$, or a class of scalar field potential $V_s(a)$, or a class of scalar field potential in terms of $\phi$, i.e. $V_s(\phi)$ (provided $\phi(a)$ is invertible), that corresponds to the naked singularity as an end state. Similarly, in the case of an SH vector field collapse, one has a class of vector field component function $A(a)$, or a class of vector field potential $V_v(a)$, or a class of vector field potential in terms of $B=g(\Tilde{A},\Tilde{A})$, i.e. $V_s(B)$ (provided $B(a)$ is invertible), that corresponds to the naked singularity as an end state.}

    \item A naked singularity formed due to gravitational collapse may or may not be relevant if they are not gravitationally strong in the sense of Tipler
    \cite{Tipler}. \textcolor{black}{Here, we show a class of $\omega(a)$ that satisfies the inequalities (\ref{strongnaked}) that corresponds to the formation of a strong curvature naked singularity. Using arguments similar to point no. $2$ of this section, we have equivalently shown a class of scalar field potential (in case of scalar field collapse) and a class of vector field potential (in case of vector field collapse) that corresponds to a strong curvature naked singularity.}
    
    \item For the sake of completion, we study the global spacetime, consisting of the interior collapsing scalar/vector field and the exterior generalized Vaidya spacetime. The smooth matching demands a restriction on the free function, that is, the generalized Vaidya mass function, in terms of the property of the interior collapsing scalar/vector field. We have fulfilled this demand by deriving the expression of the generalized Vaidya mass in terms of the equation of state parameter of the interior collapsing field in Eq.(\ref{GV1}).

\end{enumerate}




\begin{thebibliography}{}
\bibitem{Oppenheimer1939} J. R. Oppenheimer and H. Snyder, \href{https://journals.aps.org/pr/issues/56/5}{Phys. Rev. Journals Archive \textbf{ 56}, 455 (1939).}

\bibitem{Datt} S. Datt, Zs. f. Phys. \textbf{108} 314 (1938).

\bibitem{Wainwright_2005} G.F.R. Ellis, S.T.C. Siklos and J. Wainwrighit, in Dynamical systems in cosmology, Eds. J. Wainwright and G.F.R. Ellis, (Cambridge University Press, Cambridge, England, 1997).

\bibitem{penrose2} R. Penrose, Riv. Nuovo Cimento Soc. Ital. Fis. \textbf{1}, 252 (1969).

\bibitem{Geroch_70} R. Geroch, Journal of Mathematical Physics, \textbf{11}, 2, 437-449 (1970).

\bibitem{hawking} S. W. Hawking and G. F. R. Ellis, The large scale structure of spacetime, Cambridge University Press (1973).

\bibitem{Joshi_Global} P. S. Joshi, Global Aspects in Gravitation and Cosmology
(Clendron Press, Oxford, 1993).
 
\bibitem{geroch} R. Geroch and G. Horowitz, `Global structure of spacetimes', in General Relativity: An Einstein Centenary Survey, eds S. W. Hawking and W. Israel, Cambridge: Cambridge University Press (1979).

\bibitem{hawking2} S. W. Hawking and W.Israel, `An introductory survey', in General Relativity: An Einstein Centenary Survey, eds S. W. Hawking and W. Israel. Cambridge: Cambridge University Press (1979).

\bibitem{penrose3} R. Penrose, `Singularities and time asymmetry', in General Relativity: An Einstein Centenary Survey, eds S. W. Hawking and W. Israel. Cambridge: Cambridge University Press (1979).

\bibitem{penrose} R. Penrose, Phys. Rev. Lett. \textbf{14}, 57 (1965).

\bibitem{Joshi2011} P. S. Joshi and D. Malafarina, \href{https://journals.aps.org/prd/abstract/10.1103/PhysRevD.83.024009}{Phys. Rev. D \textbf{83}, 024009 (2011).}

\bibitem{psjoshi2}  P. S. Joshi, \textit{Gravitational Collapse, and Spacetime Singularities}, (Cambridge University Press, Cambridge, England, 2007).

\bibitem{Mosani2020} K. Mosani, D. Dey, P. S. Joshi, Phys. Rev. D \textbf{102}, 044037.

\bibitem{Christodoulou_94} D. Christodoulou,
Annals of Mathematics Annals of Mathematics, \textbf{140}, 607 (1994).

\bibitem{Christodoulou_99} D. Christodoulou, Annals of Mathematics, \textbf{149}, 183 (1999).

\bibitem{goswami2007} R. Goswami and P. S. Joshi, Modern Physics Letters A, \textbf{22}, 01, pp. 65-74 (2007).

\bibitem{Mosani2022} Karim Mosani, Dipanjan Dey, Kaushik Bhattacharya
and Pankaj S. Joshi, \href{https://journals.aps.org/prd/abstract/10.1103/PhysRevD.105.064048}{Phys. Rev. D \textbf{105}, 064048 (2022).}

\bibitem{Wang_99} A. Wang and Y. Wu, Gen. Relativ. Gravit. \textbf{31}, 107 (1999).

\bibitem{Tipler} F. J. Tipler, \href{https://www.sciencedirect.com/science/article/abs/pii/0375960177905084?via%3Dihub}{Phys. Lett. \textbf{64}A, 8 (1977)}.

\bibitem{Klorak} C. J. S. Clarke and A. Krolak, \href{https://www.sciencedirect.com/science/article/pii/0393044085900129?via%3Dihub}{J. Geom. Phys. \textbf{2}, 127 (1985)}.

\bibitem{Eardley_1976} D. M. Eardley, in 'Gravitation in Astrophysics', ed. B. Carter and J. B. Hartle (Plenum, New York, 1987).

\bibitem{Mosani2020a} Karim Mosani, Dipanjan Dey
and Pankaj S. Joshi, \href{https://journals.aps.org/prd/abstract/10.1103/PhysRevD.101.044052}{Phys. Rev. D \textbf{ 101}, 044052 (2020).}

\bibitem{Christodoulou}Demetrios Christodoulou,
\href{https://link.springer.com/article/10.1007/BF01205930}{Commun. Math. Phys. \textbf{105}, 337-361 (1986)}.

\bibitem{Virbhadra}K. S. Virbhadra, S. Jhingan and P. S. Joshi, \href{https://www.worldscientific.com/doi/abs/10.1142/S0218271897000200}{International Journal of Modern Physics D \textbf{06}, 357-361 (1997)}.

\bibitem{Garfinkle}David Garfinkle, Robert Mann, and Chris Vuille \href{https://journals.aps.org/prd/abstract/10.1103/PhysRevD.68.064015}{Phys. Rev. D \textbf{68}, 064015 (2003)}.

\bibitem{Poisson:2009pwt}
E.~Poisson,
``A Relativist's Toolkit: The Mathematics of Black-Hole Mechanics,''
\href{https://doi.org/10.1017/CBO9780511606601}{Cambridge University Press, (2009).}

\end{thebibliography}
\end{document}